\documentclass[12pt,a4paper]{article}

\usepackage[utf8]{inputenc}
\usepackage{amsmath}
\usepackage{amsfonts}
\usepackage{amssymb}
\usepackage{bbold}
\usepackage[dvips,letterpaper,margin=1.00in,bottom=1.00in]{geometry}
\usepackage{mathrsfs}
\usepackage{graphicx}
\usepackage[T1]{fontenc}
\usepackage[utf8]{inputenc}
\usepackage{authblk}
\usepackage{scrextend}
\usepackage{footmisc}
\usepackage{indentfirst}
\usepackage{physics}
\usepackage[english]{babel}
\usepackage{pdflscape}
\usepackage[noadjust]{cite}
\usepackage{setspace}
\usepackage[toc,page]{appendix}

\title{Sections and Chapters}
\usepackage{cite}
\title{Analytic Solutions of Brans-Dicke Cosmology: Early Inflation and Late Time Accelerated Expansion }

\author{Medine Ildes$^{1,}$\footnote{e-mail:medine.ildes@boun.edu.tr} and Metin Arik$^{1,}$\footnote{e-mail:medtin.arik@boun.edu.tr}}

\affil{$^{1}$Department of Physics, Bogazici University, Bebek, Istanbul, Turkiye}

\begin{document}

\maketitle

\begin{abstract}
\indent We investigate the most general exact solutions of Brans-Dicke cosmology by choosing the scale factor $"a"$ as the new independent variable. It is shown that a set of three field equations can be reduced to a constraint equation and a first order linear differential equation. Thus this new set of equations is solvable when one supplies one of the following pairs of functions: ($\phi(a)$, $\rho(a)$), ($\phi(a)$, $V(a)$)
or ($\phi(a)$, $H(a)$). A universe with a single component energy-matter density is studied. It is seen that when there is no constant energy density, the Hubble function still contains a constant term which causes exponential expansion. This constant is expressed in terms of the initial values of the universe. An early universe and the present universe with dark energy are studied. In addition late time accelerated expansion is also explained with cosmic domain walls. If we take Brans-Dicke parameter $\omega > 4\times 10^{4}$ formulas of the Hubble function reduce to solutions of $\Lambda CDM$ cosmology. Therefore comparison of our results with recent observations of type Ia supernovae indicates that eighty-nine percent of present universe may consist of domain walls while rest is matter.\\
\end{abstract}

\section{Introduction}
\indent Since Einstein's General Theory of Relativity, many new modified theories have been introduced to extend it. Dirac proposed that Newton's constant may be varying with time \cite{dirac1938new}, after influential ideas were developed by Weyl \cite{weyl1917gravitationstheorie,weyl1919neue} and Eddington \cite{eddington1931preliminary}. Spacetime dependent gravitational constant can be described according to Jordan's study in \cite{jordan1955schwerkraft}. Thus his work has been accepted as the origin of the scalar-tensor theory.\\
\indent A new approach to general relativity which is based on Mach's principle was developed by Brans and Dicke in 1961 \cite{brans1961mach}. They took effective gravitational constant to be proportional to the reciprocal of a scalar field. This is called as nonminimal coupling of scalar field. More complicated nonminimal coupling terms were porposed in \cite{futamase1989chaotic,salopek1989designing,fakir1992cosmological,fakir1990improvement,makino1991density,kaiser1995primordial,mukaigawa1998finite,komatsu1999complete,
 linde2011observational}.\\
\indent Two different revolutionary steps have been taken in cosmology since 1980. First one was the construction of inflationary cosmology by A.H. Guth and A.D.Linde. It had been proposed as a solution for problems of the standard model of cosmology which are  the flatness problem, the horizon problem and the monopole problem \cite{guth1981inflationary,linde1982new}. Second one was the observational evidence of accelerated expansion of the present universe  \cite{riess1998observational,perlmutter1999measurements,tonry2003cosmological}. Standard cosmology explains this behaviour by contribution of dark energy ($\sim 68\%$), dark matter ($\sim 27\%$) and baryonic matter ($\sim 5\%$) to the total density parameter. Nature of dark energy and dark matter are still unknown.\\
\indent In the last decades cosmologists established several modified theories of gravity to investigate these two concepts \cite{bassett2006inflation,baumann2009tasi,martin2014encyclopaedia,lyth1999particle,linder1990particle,fujii2003scalar,faraoni2004cosmology,lyth2009primordial,asselmeyer2016frontier,bailin2017cosmology}. All of these theories have different field equations which govern the dynamics of the universe. Thus all these theories have different physical conclusions. \\
\indent In recent years searching for exact solutions of the field equations of scalar tensor theories has been taken more attention of scientists \cite{rasouli2016exact,beesham2015exact,belinchon2016exact,makarenko2012exact
,massaeli2017general,pintus2018mathematical}. In addition in 1990 it has been shown that there are potentials $V(\phi)$ leading to desired behaviour for the scale factor without the use of 'slow-roll approximation' usually assumed in inflationary models \cite{ellis1991exact}.\\
\indent In this text we pick up Brans-Dicke cosmology and we focus on it. Up to now some exact solutions for specific cases has been presented. While for a radiation filled universe exact solution was given in \cite{morganstern1971exact}, parametric solution for the case $p=\epsilon \rho$ was given in \cite{morganstern1971exact2}. For Bianchi type universes with arbitrary barotropic perfect fluid exact solution were found \cite{chauvet1986exact}. Parametric exact solutions for potential free universe filled with ordinary matter were given in \cite{capozziello1996exact}. On the other hand Noether's theorem was used to determine conservation laws in Brans-Dicke cosmology which includes two scalar fields and exact solutions were found \cite{paliathanasis2019conservation}. An exact solutions of modified Brans-Dicke theory was studied in \cite{mukherjee2019exact}. \\ 
\indent The major purpose of this article is to solve field equations of Brans-Dicke cosmology analytically. First, we present field equations in section 2. Then we introduce our mathematical techniques in section 3 and we rewrite the field equations with the new independent variable $"a"$. This set of equations can be reduced to a constraint equation and a first order differential equation. Exact solutions are found in section 4 for given $\phi(a)$ and $\rho(a)$, in section 5 for given  $\phi(a)$ and $V(a)$, in section 6 for given $\phi(a)$ and $H(a)$. A universe consisting of single component energy-matter density is investigated in section 7. The early epoch of the universe is studied in section 8. We present a solution for a universe which contains dark energy and matter in section 9.1 and we present a solution for a universe which consists of cosmic domain walls and matter in section 9.2. Comparison of both models with observation is illustrated by a plot. Discussion is given in section 10 and we summarize our results in conclusion. In this study we choose $\omega>4\times 10^{4}$ to be compatible with results of Einstein telescope \cite{zhang2017testing} and time delay experiments \cite{alsing2012gravitational}.\\
\section{Field Equations}
\indent In this study we will use FLWR metric given by,
\begin{align}
ds^{2}=dt^{2}-a^{2}(t)[\frac{dr^{2}}{1-kr^{2}}+r^{2}(d\theta ^{2}+\sin^{2}\theta d\varphi ^{2})].
\end{align}
\\where $k$ is the curvature parameter, $a(t)=\dfrac{R(t)}{R(t_{0})}$ is the normalized scale factor of the universe. Here $r$ has dimension of $lenght$, $a(t)$ is dimensionless and $k$ has diemsion of $lenght^{-2}$. In BDJT the Lagrange density can be written as \cite{jordan1938empirischen,jordan1947erweiterung,jordan1959gegenwartigen,thiry1948geometrie,brans1961mach} 
\begin{align}
\mathcal{L} &=\left[-\Phi R+ \omega\dfrac{1}{\Phi}g^{\mu \nu}\partial_{\mu}{\Phi} \partial_{\nu}{\Phi}-V(\Phi)+L_{M}\right]\sqrt{-g}, \\
&=\left[-\frac{1}{8\omega}\phi ^{2}R+\frac{1}{2}g^{\mu \nu}\partial_{\mu}{\phi} \partial_{\nu}{\phi}-V(\phi)+L_{M}\right]\sqrt{-g},
\end{align}
\\where we will call $\phi$ as the Jordan scalar field and $\Phi$ as the Brans-Dicke scalar field, the two being related by
\begin{align}
\Phi=\frac{1}{8\omega}\phi ^{2},
\end{align}
\\where $\omega$ is the dimensionless Brans-Dicke parameter, $R$ is the Ricci scalar and $L_{M}$ represents the contribution due to matter fields. We use the metric signature $(+,-,-,-)$ and units with $\hbar=1$, $c=1$. We prefer to use the Jordan scalar for the scalar field since in flat spacetime the Lagrangian then becomes the standard Lagrangian for a scalar field
\begin{align}
\mathcal{L} =\frac{1}{2}\partial_{\mu}{\phi} \partial^{\mu}{\phi}-V(\phi).
\end{align}
\indent The homogeneous and isotropic cosmological field equations obtained from this Lagrangian density have already been calculated for a potential $V(\phi )=\dfrac{1}{2}m^{2}\phi ^{2}$ by Arik, Calik and Sheftel in \cite{arik2005primordial,arik2006can,arik2008friedmann}. Thus their general form for a potential $V(\phi )$ are given by,
\begin{gather}
\frac{3}{4\omega}\phi ^{2}\left(\frac{\dot{a}^2}{a^2}+\frac{k}{a^{2}}\right)-\frac{1}{2}\dot{\phi }^{2}-V(\phi )+\frac{3}{2\omega}\frac{\dot{a}}{a}\dot{\phi }\phi=\rho_{m}, \\
\frac{-1}{4\omega}\phi ^{2}\left(2\frac{\ddot{a}}{a}+\frac{\dot{a}^2}{a^2}+\frac{k}{a^{2}}\right)-\frac{1}{\omega}\frac{\dot{a}}{a}\dot{\phi }\phi-\frac{1}{2\omega}\ddot{\phi }\phi-\left(\frac{1}{2}+\frac{1}{2\omega}\right)\dot{\phi }^{2}+V(\phi )=p_{m},\\
\ddot{\phi }+3\frac{\dot{a}}{a}\dot{\phi }+\dv{V(\phi )}{\phi }-\frac{3}{2\omega }\left(\frac{\ddot{a}}{a}+\frac{\dot{a}^2}{a^2}+\frac{k}{a^{2}}\right)\phi =0.
\end{gather}
\section{Mathematical Techniques}
\indent It is apparent that the independent variable time $t$ is not seen explicitly in our set of differential equations given by (6-8). For this reason one can choose a new variable in terms of old dependent variables. Hence the order of differential equation can be reduced by one \cite{rainville1989elementary}. Our choice is $\dot{a}=aH(a)$. Then $a$ becomes new independent variable and field equations transform into the following form\\
\begin{gather}
\frac{3}{4\omega}\phi^{2}\left(H^{2}+\dfrac{k}{a^{2}}\right)-\dfrac{1}{2}\left(\phi^{'}H a\right)^{2}-V(a)+\frac{3}{2\omega}\phi \phi^{'} H^{2}a=\rho_{m}  \\
-\frac{1}{4\omega}\phi^{2}\left(3H^{2}+2H^{'}H a+\dfrac{k}{a^{2}}\right)-\dfrac{3}{2\omega}\phi^{'}\phi H^{2}a-\dfrac{1}{2\omega}\left(\phi^{''}H^{2}a^{2}+\phi^{'}H^{'}H a^{2}\right)\phi\nonumber \\-\left(\dfrac{1}{2}+\dfrac{1}{2\omega}\right)\left(\phi^{'}H a\right)^{2}+V(a)=p_{m},\\
\phi^{''}H^{2}a^{2}+\phi^{'}H^{'}H a^{2}+4\phi^{'}H^{2}a+\dfrac{V^{'}}{\phi^{'}}-\dfrac{3}{2\omega}\left(2H^{2}+H^{'}H a+\frac{k}{a^{2}}\right)\phi=0,
\end{gather}
\\where the prime denotes $\dfrac{d}{da}$ (Details are given in appendix A). We will name (9) as the energy density equation, (10) as the pressure equation and (11) as the $\phi$ equation. \\
\indent These three equations are connected by the perfect fluid continuity equation,
\begin{gather}
\dot{\rho}(a)+3\dfrac{\dot{a}}{a}(\rho(a)+p(a))=0, \\
[\rho^{'}(a)a+3(\rho(a)+p(a))]H(a)=0.
\end{gather}
\\In other words when we plug energy density and pressure which are given by (9-10) in (13) we end up with $-a\phi^{'}(a)H(a)\times eq.(11)$. This shows that we have two independent equations and four unknown functions: $H(a)$, $V(a)$, $\phi(a)$ and $\rho(a)$ (one can choose $p(a)$ instead of $\rho(a)$).\\
\indent In our calculations we will use the energy density equation and the $\phi$ equation. When we plug solutions in to the pressure equation, $\rho(a)$ and the resulting $p(a)$ will satisfy the perfect fluid equation of state.\\
\indent It is seen that energy density equation is a first order variable coefficient and nonlinear differential equation for $\phi(a)$. In addition this is not a type of Bernoulli equation which is analytically solvable. In addition (11) is second order variable coefficient nonlinear differential equation for $\phi(a)$. Analytic solutions for both of them have not been established yet. To be able to solve them, $\phi(a)$ should be given. Then we will be left with $\rho(a)$, $V(a)$ and $H(a)$. Since (9) does not contain any derivative of last three functions it is just a constraint. On the other hand first derivatives of $H(a)$ and $V(a)$ are seen in (11). Thus our system consist of a constraint equation and a first order differential equation. One of the following functions; $H(a)$, $V(a)$ and $\phi(a)$ is still free. One of them should be accompanied with $\phi(a)$ as a given function. Three combinations exist; $\rho(a)$ and $\phi(a)$, $V(a)$ and $\phi(a)$, $H(a)$ and $\phi(a)$. We will solve our system for these three sets separately.  \\
\section{Solution for given $\rho(a)$ and $\phi(a)$}
\indent One can solve the constraint equation (9) and the first order differential equation (11) by different methods:
\begin{itemize}
\item Method 1) First, find $V(a)$ from the constraint equation then substitute it in the differential equation and solve it for $H(a)$.
\item Method 2) First, solve the differential equation for $V(a)$ then substitute it in the constraint equation and solve it for $H(a)$.
\item Method 3) First, find $H(a)$ from the constraint equation then substitute it in the differential equation and solve it for $V(a)$.
\item Mehod 4) First, solve the differential equation for $H(a)$ then substitute it in the constraint equation and solve it for $V(a)$.
\end{itemize}
\indent Results are different representations of the same solution. However they may have different singular cases which require special attention. To reach complete set of solutions we perform all four methods. Moreover, most of the solutions include integrals of different functions and sometimes integration techniques can be inadequate . For this reason one of the form of solutions can be superior to others. We would like to remind that the $\rho$ equation is equivalent to the constraint equation and the first order differential equation is equivalent to the $\phi$ equation. We will use both names.\\
\subsection{Method 1}
\subsubsection{Nonsingular case}
\indent Firstly we find $V(a)$ from the constraint equation
\begin{align}
V(a)=\dfrac{3}{4\omega}\phi^{2}(a)(H^{2}(a)+\dfrac{k}{a^{2}})-\dfrac{1}{2}(\phi^{'}(a)H(a)a)^{2}+\dfrac{3}{2\omega}\phi \phi^{'}H^{2}a+\rho(a).
\end{align}
\\Then the $\phi$ equation becomes
\begin{align}
H^{'}(a)+\dfrac{a[(1+2\omega)\phi^{'2}(a)+\phi(a) \phi^{''}(a)]}{\phi(a) (\phi(a)+a\phi^{'}(a))}H(a)=\dfrac{2\omega\rho^{'}a^{3}+3k\phi^{2}(a)}{3a^{3}\phi(a) (\phi(a)+a\phi^{'}(a))H(a)},
\end{align}
\\which is the Bernoulli equation (Details are given in appendix C). Thus we apply change of variable $\gamma(a)=H^{2}(a)$ and we obtain
\begin{align}
\gamma^{'}(a)+\dfrac{2a[(1+2\omega)\phi^{'2}(a)+\phi(a) \phi^{''}(a)]}{\phi(a) (\phi(a)+a\phi^{'}(a))}\gamma(a)=\dfrac{2(2\omega\rho^{'}a^{3}+3k\phi^{2}(a))}{3a^{3}\phi(a) (\phi(a)+a\phi^{'}(a))}.
\end{align}
\\Its solution is found as
\begin{gather}
\gamma(a)=\dfrac{exp\Big[-\int_{a_{in}}^{a}P(a^{'})da^{'}\Big]}{\Big[\phi (\phi+a\phi^{'})\Big]^{2}}\Big\{\int_{a_{in}}^{a}exp\Big[\int_{a_{in}}^{a^{'}}P(a^{''})da^{''}\Big]Q(a^{'})da^{'}+\tilde{\gamma}(a_{in})\Big\},  \\ \nonumber \\
P(a)=\dfrac{2[2\omega a\phi^{'2}(a)-3\phi(a) \phi^{'}(a)]}{\phi(a) (\phi(a)+a\phi^{'}(a))}, \hspace{20pt}Q(a)=\dfrac{2(2\omega\rho^{'}a^{3}+3k\phi^{2}(a))\Big[\phi(a) (\phi(a)+a\phi^{'}(a))\Big]}{3a^{3}}, \\ \nonumber\\ \tilde{\gamma}(a_{in})=\gamma(a_{in})\Big[\phi(a_{in})\Big(\phi(a_{in})+a_{in}\phi^{'}(a_{in})\Big)\Big]^{2},
\end{gather}
\\(Details are given in appendix B and D). Hence
\begin{align}
H(a)=\sqrt{\gamma(a)},
\end{align}
\\where $\gamma(a)$ is given by (17). One should choose $\phi(a)$ such that $\gamma(a)$ is positive. Thus $H(a)$  will be a real function. However if one goes further by assigning specific form to the Hubble function and tries to solve (17) for $\phi(a)$, she will end up with a second order, variable coefficient, nonlinear differential equation for whose solution there is no known method. For this reason we will continue with trial $\phi(a)$ functions in our examples.\\
\subsubsection{Singular case}
\indent The denominator of the function $P(a)$ which is used to formulate the Hubble function becomes singular when $\phi(a)=F/a$. Hence this case needs special attention. $V(a)$ is obtained form the $\rho$ equation,
\begin{align}
V(a)=-\rho(a)+\dfrac{F^{2}\Big[3k-(3+2\omega)a^{2}H^{2}(a)\Big]}{4\omega a^{4}}.
\end{align}
\\Then the $\phi$ equation turns into the following form
\begin{align}
\dfrac{3F^{2}\Big[k-(3+2\omega)a^{2}H^{2}(a)\Big]+2\omega a^{5}\rho^{'}(a)}{2F\omega a^{3}}=0.
\end{align}
\\The Hubble function is found as
\begin{align}
H(a)=\sqrt{\dfrac{3F^{2}k+2\omega a^{5}\rho^{'}(a)}{3(3+2\omega)F^{2}a^{2}}}.
\end{align}
\\Hence the final form of potential is given by
\begin{align}
V(a)=-\rho(a)-\dfrac{a}{6}\rho^{'}(a)+\dfrac{kF^{2}}{2\omega a^{4}}.
\end{align}
\\If one inserts the energy density of an ordinary matters which is proportional to $1/a^{n}$ results are
\begin{gather}
H=\sqrt{\dfrac{3F^{2}k-2n\omega\rho_{n}a^{4-n}}{3(3+2\omega)F^{2}a^{2}}},\\
V(a)=\dfrac{F^{2}k}{2\omega a^{4}}+\dfrac{(n-6)\rho_{n}}{a^{n}},\\
V(\phi)=\dfrac{F^{2}k}{2\omega}\phi^{4}+(n-6)\rho_{n}\phi^{n}.
\end{gather}
\\ Except the radiation dominated universe and dark energy dominated universe all these matters results in an imaginary form of the Hubble function and negative potential values as easily recognized from (25-26). In addition radiation dominated universe corresponds to $V=\lambda\phi^{4}/4$ which has been studied in \cite{arik2018inflation}.\\
\subsection{Method 2}
\subsubsection{Non-singular Case}
\indent In the second method we start with the $\phi$ equation. By using it $V(a)$ is found as
\begin{align}
V(a)=-\int_{a_{in}}^{a}\dfrac{\phi^{'}\Big\{-3\phi\Big[k+a^{'2}H(2H+a^{'}H^{'})\Big]+2a^{'3}\omega H\Big[(4H+a^{'}H^{'})\phi^{'}+a^{'}H\phi^{''}\Big]\Big\}}{2a^{'2}\omega}da^{'}+V(a_{in}).
\end{align}
\\Then we insert $V(a)$ in $\rho$ equation and we take the derivative of both sides with respect to $a$. We obtain the following equation
\begin{align}
H^{'}(a)+\dfrac{a[(1+2\omega)\phi^{'2}(a)+\phi(a) \phi^{''}(a)]}{\phi(a) (\phi(a)+a\phi^{'}(a))}H(a)=\dfrac{2\omega\rho^{'}(a)a^{3}+3k\phi^{2}(a)}{3a^{3}\phi(a) (\phi(a)+a\phi^{'}(a))H(a)}. 
\end{align}
\\This equation is equivalent to (15). Its solution has been already presented  in (17-20). Although forms of potentials which are given by  (24) and (28) are different, one can show that they are equal to each other up to a constant. \\
\indent We have just shown that results of method 1 and method 2 are the same, thus there is no need to repeat calculations for singular case. In this method we end up with two integration constants. One of them should be chosen such that when final forms of formulas are inserted in the constraint equation it must be satisfied.
\subsection{Method 3}
\subsubsection{Non Singular Case}
\indent First we find $H(a)$ from the $\rho$ equation
\begin{align}
H(a)=\sqrt{\dfrac{-3k\phi^{2}(a)+4\omega a^{2}\Big[\rho(a)+V(a)\Big]}{a^{2}\Big[3\phi^{2}(a)+6a\phi(a)\phi^{'}(a)-2\omega a^{2}\phi^{'2}\Big]}}.
\end{align}
\\Then the $\phi$ equation turns into the following form
\begin{align}
V^{'}(a)+P(a)V(a)=Q(a),
\end{align}
\\where
\begin{gather}
P(a)=\Big\{2 \phi^{'}(a)\Big[-6 \phi^{3}(a)+(3+14 \omega)a^{2} \phi(a) \phi^{'2}(a)-2\omega(1+2\omega)a^{3} \phi^{'3}(a)+a
\phi^{2}(a)((-6+8 \omega) \phi^{'}(a)\nonumber \\
+(3+2 \omega)a \phi^{''}(a))\Big]\Big\}/\Big\{\phi(a)(\phi(a)+a \phi^{'}(a))\Big[3 \phi^{2}(a)+6 a \phi(a) \phi^{'}(a)-2\omega a^{2}\phi^{'2}(a)\Big]\Big\},
\end{gather}
\\and
\begin{gather}
Q(a)=-\Big\{\phi^{'}(a) \Big[-36\omega ^{2}a^{4} \phi(a)\rho^{'}(a) \phi^{'2}(a)+8\omega ^{3}a^{5}\rho^{'}(a) \phi^{'3}(a)-12\omega a^{2} \phi^{2}(a) \phi^{'}(a)((-3+\omega)a\rho^{'}(a)\nonumber \\+k
(3+2\omega) \phi^{'2}(a))+9 a \phi^{3}(a)(2\omega a\rho^{'}(a)+3k(3+2 \omega) \phi^{'2}(a))+9k(3+2\omega) \phi^{4}(a)(3 \phi^{'}(a)+a \phi^{''})\nonumber \\+12\omega a\rho(a)(6 \phi^{3}(a)-(3+14\omega)a^{2}\phi(a) \phi^{'2}(a)+2\omega(1+2\omega)a^{3}\phi^{'3}(a)-a\phi^{2}(a)((-6+8 \omega) \phi^{'}(a)\nonumber \\+(3+2\omega)a\phi^{''}(a)))\Big]\Big\}/\Big\{6\omega a\phi(a)(\phi(a)+a \phi^{'}(a)) \Big[-3 \phi^{2}(a)-6 a \phi(a) \phi^{'}(a)+2\omega a^{2}\phi^{'2}(a)\Big]\Big\}.
\end{gather}
\\Hence potential is
\begin{align}
V(a)=exp\Big(-\int_{a_{in}}^{a}P(a^{'})da^{'}\Big)\Big\{\int_{a_{in}}^{a}exp\Big(\int_{a_{in}}^{a^{'}}P(a^{''})da^{''}\Big)Q(a^{'})da^{'}+V(a_{in})\Big\}.
\end{align}
\subsubsection{Singular Cases}
\indent The solution for method 3 has three singular cases;\\ \\
I) $\phi(a)=0$, \\ \\
II)$\phi(a)+a\phi^{'}(a)=0$,\\ \\
III)$-3 \phi^{2}(a)-6 a \phi(a) \phi^{'}(a)+2\omega a^{2}\phi^{'2}(a)=0$. \\
\\For all of them both functions, $P(a)$ and $Q(a)$ which are used to formulate the potential become undefined. \\
\indent First case is not acceptable because Brans-Dicke theory collapses for $\phi(a)=0$. Second case has been already investigated in Method 1. We are left with the last one. We write $\phi(a)=Fexp(\int \alpha(a)da)$. Then singularity III becomes
\begin{align}
2\omega a^{2}\alpha^{2}-6a\alpha-3=0.
\end{align}
Thus
\begin{align}
\alpha(a)=\dfrac{3\pm \sqrt{9+6\omega}}{2\omega a},\hspace{20pt}
\phi(a)=Fa^{(3\pm\sqrt{9+6\omega})/(2\omega)}.
\end{align}
\\Then the potential is found from the $\rho$ equation
\begin{align}
V(a)=\dfrac{3F^{2}k}{4\omega}a^{(3-2\omega\pm \sqrt{9+6\omega})/\omega}-\rho(a).
\end{align}
\\After writing $H(a)=\sqrt{\gamma (a)}$, $\phi$ equation becomes
\begin{align}
\gamma^{'}(a)+P(a)\gamma (a)=Q(a).
\end{align}
\\For $\alpha(a)=\dfrac{3+\sqrt{9+6\omega}}{2\omega a}$,
\begin{gather}
P(a)=\dfrac{2(3+3\omega+\sqrt{9+6\omega})}{\omega a},\\ \nonumber \\
Q(a)=\dfrac{4\omega a^{-(3+3\omega+\sqrt{9+6\omega})/\omega}}{3F^{2}(3+2\omega+\sqrt{9+6\omega})}\Big[3F^{2}ka^{(3+\sqrt{9+6\omega})/\omega}+2\omega a^{3}\rho^{'}(a)\Big].
\end{gather}
\\Then
\begin{gather}
\gamma(a)=\dfrac{1}{a^{2(3+3\omega+\sqrt{9+6\omega})/\omega}}\Big\{b\int_{a_{in}}^{a}a^{'(3+3\omega+\sqrt{9+6\omega})/\omega}\Big[3F^{2}ka^{'(3+\sqrt{9+6\omega})/\omega}+2 \omega a^{'3}\rho^{'}\Big]da^{'}+\tilde{\gamma}(a_{in})\Big\},\\
b=\dfrac{4\omega}{3F^{2}(3+2\omega+\sqrt{9+6\omega})}, \hspace{25pt}\tilde{\gamma}(a_{in})=a_{in}^{2(3+3\omega+\sqrt{9+6\omega})/\omega}\gamma(a_{in}) 
\end{gather}
\\For $\alpha(a)=\dfrac{3-\sqrt{9+6\omega}}{2\omega a}$,
\begin{gather}
P(a)=\dfrac{2(3+3\omega-\sqrt{9+6\omega})}{\omega a},\\ \nonumber \\
Q(a)=\dfrac{4\omega a^{-3(1+\omega)/\omega}}{3F^{2}(3+2\omega-\sqrt{9+6\omega})}\Big[3F^{2}ka^{3/\omega}+2\omega a^{(3\omega+\sqrt{9+6\omega})/\omega}\rho^{'}(a)\Big].
\end{gather}
\\Then
\begin{gather}
\gamma(a)=\dfrac{1}{a^{2(3+3\omega-\sqrt{9+6\omega})/\omega}}\Big\{b\int_{a_{in}}^{a}a^{'(3+3\omega-2\sqrt{9+6\omega})/\omega}\Big[3F^{2}ka^{'3/\omega}+2\omega a^{'(3\omega+\sqrt{9+6\omega})/\omega}\rho^{'}\Big]da^{'}+\tilde{\gamma}(a_{in})\Big\},\\
b=\dfrac{4\omega}{3F^{2}(3+2\omega-\sqrt{9+6\omega})}, \hspace{25pt}\tilde{\gamma}(a_{in})=a_{in}^{2(3+3\omega-\sqrt{9+6\omega})/\omega}\gamma(a_{in}) 
\end{gather}
\\At the next step we examine nature of the Hubble function. First, we take 
\begin{align}
\phi(a)=Fa^{(3+\sqrt{9+6\omega})/(2\omega)}, \hspace{20pt} \rho=\dfrac{\rho_{n}}{a^{n}}.
\end{align}
\\Hubble function and the potential are found as
\begin{gather}
H(a)=\dfrac{1}{a^{(3+3\omega+\sqrt{9+6\omega})/\omega}}\Big\{\dfrac{2\omega^{2}a^{(3-(n-4)\omega+\sqrt{9+6\omega})/\omega}}{3F^{2}(3+2\omega+\sqrt{9+6\omega})^{2}[3+(6-n)\omega+\sqrt{9+6\omega}]}\nonumber \\
\Big[-4(3+2\omega+\sqrt{9+6\omega})n\omega\rho_{n}a^{2}+3F^{2}k\Big(3+(6-n)\omega+\sqrt{9+6\omega}\Big)a^{(3+n\omega+\sqrt{9+6\omega})/\omega}\Big]+\tilde{\gamma}(a_{in})\Big\}^{1/2}, \\
\tilde{\gamma}(a_{in})=a_{in}^{2(3+3\omega+\sqrt{9+6\omega})/\omega}\gamma(a_{in})-\Big\{\dfrac{2\omega^{2}a^{(3-(n-4)\omega+\sqrt{9+6\omega})/\omega}_{in}}{3F^{2}(3+2\omega+\sqrt{9+6\omega})^{2}[3+(6-n)\omega+\sqrt{9+6\omega}]}\nonumber \\
\Big[-4(3+2\omega+\sqrt{9+6\omega})n\omega\rho_{n}a^{2}_{in}+3F^{2}k\Big(3+(6-n)\omega+\sqrt{9+6\omega}\Big)a^{(3+n\omega+\sqrt{9+6\omega})/\omega}_{in}\Big]\Big\},\\
V(a)=-\dfrac{\rho_{n}}{a^{n}}+\dfrac{3F^{2}k}{4\omega a^{(2\omega-3-\sqrt{9+6\omega})/\omega}}.
\end{gather}
\\The term containing energy density in $H(a)$ has a negative coefficient for $0< n< 6$ and $\omega\gg 1$.\\
\indent Then we take
\begin{align}
\phi(a)=Fa^{(3-\sqrt{9+6\omega})/(2\omega)}, \hspace{20pt} \rho=\dfrac{\rho_{n}}{a^{n}}.
\end{align}
\\Hubble function and the potential are found as
\begin{gather}
H(a)=\dfrac{1}{a^{(3+3\omega-\sqrt{9+6\omega})/\omega}}\Big\{\dfrac{2\omega^{2}a^{(3-(n-4)\omega-2\sqrt{9+6\omega})/\omega}}{3F^{2}(-3-2\omega+\sqrt{9+6\omega})^{2}[-3+(n-6)\omega+\sqrt{9+6\omega}]}\nonumber \\
\Big[4(3+2\omega-\sqrt{9+6\omega})n\omega\rho_{n}a^{(2\omega+\sqrt{9+6\omega})/\omega}+3F^{2}k\Big(-3+(n-6)\omega+\sqrt{9+6\omega}\Big)a^{3/\omega+n}\Big]+\tilde{\gamma}(ain)\Big\}^{1/2}, \\
\tilde{\gamma}(a_{in})=a_{in}^{2(3+3\omega-\sqrt{9+6\omega})/\omega}\gamma(a_{in})-\Big\{\dfrac{2\omega^{2}a^{(3-(n-4)\omega-2\sqrt{9+6\omega})/\omega}_{in}}{3F^{2}(-3-2\omega+\sqrt{9+6\omega})^{2}[-3+(n-6)\omega+\sqrt{9+6\omega}]}\nonumber \\
\Big[4(3+2\omega-\sqrt{9+6\omega})n\omega\rho_{n}a^{(2\omega+\sqrt{9+6\omega})/\omega}_{in}+3F^{2}k\Big(-3+(n-6)\omega+\sqrt{9+6\omega}\Big)a^{3/\omega+n}_{in}\Big]\Big\},\\
V(a)=-\dfrac{\rho_{n}}{a^{n}}+\dfrac{3F^{2}k}{4\omega a^{(2\omega-3+\sqrt{9+6\omega})/\omega}}.
\end{gather}
\\The term containing energy density in $H(a)$ has a negative coefficient for $0< n< 6$ and $\omega\gg 1$.
\subsection{Method 4}
\indent Substitution of $H(a)=\sqrt{\gamma(a)}$ into $\phi$ equation results in the following form,
\begin{gather}
\gamma^{'}(a)+\tilde{P}(a)\gamma=\tilde{Q}(a), \\ 
\tilde{P}(a)=\dfrac{4\Big[-3\phi(a)+\omega a(4\phi^{'}(a)+a\phi^{''}(a))\Big]}{a(-3\phi(a)+2\omega a\phi^{'}(a))}, \hspace{20pt} \tilde{Q}(a)=\dfrac{-4\omega a^{2}V^{'}(a)+6k\phi(a)\phi^{'}(a)}{a^{3}\phi^{'}(a)(-3\phi(a)+2\omega a\phi^{'}(a))}. 
\end{gather}
\\Hence
\begin{gather}
\gamma(a)=exp\Big(-\int_{a_{in}}^{a}\tilde{P}(a^{'})da^{'}\Big)\Big\{\int_{a_{in}^{a}}exp\Big(\int_{a_{in}}^{a^{'}}\tilde{P}(a^{''})da^{''}\Big)\tilde{Q}(a^{'})da^{'}+\gamma(a_{in})\Big\},\\
H(a)=\sqrt{\gamma(a)}.
\end{gather}
\\After plugging $H(a)$ into $\rho$ equation we multiply both sides by
\begin{align}
\dfrac{exp\Big(\int_{a_{in}}^{a}\tilde{P}(a^{'})da^{'}\Big)}{3\phi^{2}(a)+6a\phi(a)\phi^{'}(a)-2 \omega a^{2}\phi^{'2}}.
\end{align}
\\Then we take derivative of both sides with respect $a$. The resulting equation is equivalent to
\begin{align}
V^{'}(a)+P(a)V(a)=Q(a)
\end{align} 
\\where coefficients $P(a)$ and $Q(a)$ are given by (32) and (33) respectively. Thus solution is given by (34). In this method we end up with two integration constants. One of them should be chosen such that when final forms of formulas are inserted in the constraint equation it must be satisfied.\\
\subsubsection{Singularities}
\indent This method has five singularities in its formulation. Three of them are the same singularities given in Method 3. Other two of them are seen in functions $\tilde{P}(a)$ and $\tilde{Q}(a)$\\ \\
I)$\phi^{'}(a)=0$, \\ \\
II)$-3\phi(a)+2\omega a\phi^{'}(a)=0$.\\
\\First one makes Brans-Dicke cosmology equivalent to Einstein cosmology which have been studied in \cite{ildes2022analytic}. The second one implies
\begin{align}
\phi(a)=Fa^{3/(2\omega)}.
\end{align}
\\Then $\phi$ equation becomes
\begin{align}
\dfrac{a^{-3/\omega-2}}{12F\omega^{2}}\Big\{-9F^{2}a^{3/\omega}\Big[2k\omega-(3+2\omega)a^{2}H^{2}(a)\Big]+8\omega^{3}a^{3}V^{'}(a)\Big\}=0.
\end{align}
\\The Hubble function is found easily
\begin{align}
H(a)=\dfrac{\sqrt{18F^{2}k\omega a^{3/\omega}-8\omega^{3}a^{3}V^{'}(a)}}{3\sqrt{(3+2\omega)F^{2}a^{3/\omega+2}}}.
\end{align}
\\Hence the $\rho$ equation turns into the following form
\begin{gather}
V^{'}(a)+P(a)V(a)=Q(a), \\
P(a)=\dfrac{3}{\omega a}, \hspace{20pt}Q(a)=\dfrac{3(3F^{2}ka^{3/\omega}-2\omega a^{2}\rho(a))}{2\omega^{2}a^{3}}.
\end{gather}
\\Its solution is
\begin{gather}
V(a)=\dfrac{1}{a^{3/\omega}}\Big\{\int_{a_{in}}^{a}a^{'3/\omega}Q(a^{'})da^{'}+\tilde{V}(a_{in})\Big\},\\
\tilde{V}(a_{in})=a_{in}^{3/\omega}V(a_{in}).\nonumber
\end{gather}
\\To investigate the nature of the Hubble function and the potential for usual forms of energy-matters we plug $\rho=\rho_{n}/a^{n}$. Thus we have
\begin{gather}
H(a)=\omega\Big\{\frac{2}{3(3+2\omega)}\Big[\dfrac{4a_{in}^{3/\omega}V(a_{in})}{F^{2}a^{6/\omega}}-\dfrac{3k}{(\omega-3)a^{2}}+\dfrac{4n\omega\rho_{n}}{(n\omega-3)F^{2}a^{3/\omega+n}}\Big]\Big\}^{1/2},\\
V(a)=\dfrac{\tilde{V}(a_{in})}{a^{3/\omega}}+\dfrac{3\rho_{n}}{(n\omega-3)a^{n}}+\dfrac{9F^{2}k}{4\omega(3-\omega)a^{-3/\omega+2}},\\
\tilde{V}(a_{in})=a_{in}^{3/\omega}V(a_{in})-\dfrac{3\rho_{n}}{(n\omega-3)a^{n}_{in}}-\dfrac{9F^{2}k}{4\omega (3-\omega)a^{-3/\omega+2}_{in}}.
\end{gather} 
\\Energy density related term in $H(a)$ has a positive coefficient therefore the Hubble function is always real.\\
\section{Solution for given $V(a)$ and $\phi(a)$}
\indent Special case of potential in the scalar-tensor theory is important. For this reason formulation of this case is significant. Since in general the potential is given as a function of $\phi$, we would like to remind that the potential as a function of $"a"$ is simply found as $V(a)=V(\phi (a))$.  We establish the solution for our set of the constraint equation and the first order differential equation by two following different methods:
\begin{itemize}
\item Method 1) First, solve the $\phi$ equation for $H(a)$ then plug it into the constraint equation to obtain $\rho(a)$.
\item Method 2) First, find $H(a)$ from the constraint equation then plug it in to the $\phi$ equation and solve it for $\rho(a)$.
\end{itemize} 
\subsection{Method1}
\subsubsection{Non-singular Case}
\indent We start our calculations by changing the dependent variable as $H(a)=\sqrt{\gamma (a)}$ in $\phi$ equation. Hence it turns into the following form
\begin{gather}
\gamma^{'}(a)+P(a)\gamma (a)=Q(a),\\ \nonumber \\
P(a)=\dfrac{4\Big[-3\phi(a)+\omega a(4\phi^{'}(a)+a\phi^{''}(a))\Big]}{a(-3\phi(a)+2\omega a\phi^{'}(a))},\\ \nonumber \\
Q(a)=\dfrac{-4\omega a^{2}V^{'}(a)+6k\phi(a)\phi^{'}(a)}{a^{3}\phi^{'}(a)(-3\phi(a)+2\omega a\phi^{'}(a))}.
\end{gather}
\\Its solution is given by
\begin{align}
\gamma(a)=exp\Big(-\int_{a_{in}}^{a}P(a^{'})da^{'}\Big)\Big\{\int_{a_{in}}^{a}exp\Big(\int_{a_{in}}^{a^{'}}P(a^{''})da^{''}\Big)Q(a^{'})da^{'}+\gamma(a_{in})\Big\}.
\end{align}
\\Energy density is obtained by inserting $H(a)=\sqrt{\gamma (a)}$ and $V(a)$ in the $\rho$ equation.
\subsubsection{Singular Case}
\indent When $-3\phi(a)+2\omega a\phi^{'}(a)=0$, previous solution becomes singular. This is the singularity of section 4.4. Thus $\phi(a)=Fa^{3/(2\omega)}$ and $\phi$ equation is equal to (62) so $H(a)$ is found in (63). Then energy density is found as
\begin{align}
\rho(a)=\dfrac{3F^{2}ka^{3/\omega-2}}{2\omega}-V(a)-\dfrac{\omega aV^{'}(a)}{3}.
\end{align}
\\When we inspect the results for $V=\dfrac{V_{n}}{a^{n}}$ we obtain
\begin{gather}
H(a)=\dfrac{1}{3\sqrt{3+2\omega}}\sqrt{\dfrac{18k\omega}{a^{2}}+\dfrac{8n\omega^{3}V_{n}}{F^{2}a^{3/\omega+n}}}, \\
\rho(a)=\dfrac{3F^{2}ka^{3/\omega-2}}{2\omega}+\dfrac{(n\omega-3)V_{n}}{3a^{n}}.
\end{gather}
\subsection{Method 2}
\subsubsection{Non-singular Case}
First we find $H(a)$ from the $\rho$ equation
\begin{align}
H(a)=\sqrt{\dfrac{-3k\phi^{2}(a)+4\omega a^{2}\Big[\rho(a)+V(a)\Big]}{a^{2}\Big[3\phi^{2}(a)+6a\phi(a)\phi^{'}(a)-2\omega a^{2}\phi^{'2}\Big]}}.
\end{align}
\\Then $\phi$ equation turns into the following form
\begin{align}
\rho^{'}(a)+P(a)\rho(a)=Q(a),
\end{align}
\\where
\begin{gather}
P(a)=\Big\{6\Big[6\phi^{3}(a)-(3+14\omega)a^{2}\phi(a)\phi^{'2}(a)+2(1+2\omega)\omega a^{3}\phi^{'3}(a)-a\phi^{2}(a)((-6+8\omega)\phi^{'}(a)\nonumber \\
+(3+2\omega)a\phi^{''}(a))\Big]\Big\}/\Big\{a(-3\phi(a)+2\omega a\phi^{'}(a))\Big[-3\phi^{2}(a)-6a\phi(a)\phi^{'}(a)+2\omega a^{2}\phi^{'2}(a)\Big]\Big\}, \\
Q(a)=-\Big\{3\Big[4 \omega a^{3}\phi(a)(-(3+14\omega)V(a)+\omega aV^{'}(a))\phi^{'3}(a)+8\omega^{2}(1+2\omega)a^{4}V(a)\phi^{'4}(a)\nonumber \\+3a\phi^{3}(a)\phi^{'}(a)(8\omega V(a)-6\omega aV^{'}(a)+3k(3+2\omega)\phi^{'2}(a))+4\omega a^{2}\phi^{2}(a)\phi^{'}(a)\nonumber \\ \Big(\phi^{'}(a)((6-8\omega)V(a)+(-3+\omega)aV^{'}(a)-k(3+2\omega)\phi^{'2}(a))-(3+2\omega)aV(a)\phi^{''}(a)\Big)\nonumber \\+3\phi^{4}(a)\Big(-2\omega aV^{'}(a)+k(3+2\omega)\phi^{'}(a)(3\phi^{'}(a)+a\phi^{''}(a))\Big)\Big]\Big\}/\nonumber \\ \Big\{2\omega a^{2}\phi^{'}(a)(-3\phi(a)+2\omega a\phi^{'}(a))\Big[-3\phi^{2}(a)-6a\phi(a)\phi^{'}(a)+2\omega a^{2}\phi^{'2}(a)\Big]\Big\}.
\end{gather}
\\Hence energy density is found as
\begin{align}
\rho(a)=exp\Big(-\int_{a_{in}}^{a}P(a^{'})da^{'}\Big)\Big\{\int_{a_{in}}^{a}exp\Big(\int_{a_{in}}^{a^{'}}P(a^{''})da^{''}\Big)Q(a^{'})da^{'}+\rho(a_{in})\Big\}.
\end{align}
\subsubsection{Singular Cases}
\indent This method has two singular cases for non-constant scalar field.\\
I)$-3\phi(a)+2\omega a\phi^{'}(a)=0$,\\ \\
II)$-3 \phi^{2}(a)-6 a \phi(a) \phi^{'}(a)+2\omega a^{2}\phi^{'2}(a)=0$. \\
\indent First case implies $\phi(a)=Fa^{3/(2\omega)}$. Thus the $\rho$ equation becomes
\begin{align}
\dfrac{3F^{2}a^{3/\omega-2}\Big[2k\omega+(3+2\omega)a^{2}H^{2}(a)\Big]}{8\omega^{2}}-V(a)=\rho(a).
\end{align}
\\The Hubble function is found as
\begin{align}
H(a)=\sqrt{\dfrac{\omega}{3+2\omega}}\sqrt{-\dfrac{2k}{a^{2}}+\dfrac{8\omega}{3F^{2}a^{3/\omega}}(\rho(a)+V(a))}.
\end{align}
\\Then the $\phi$ equation becomes
\begin{align}
-\dfrac{3Fk}{\omega}a^{3/(2\omega)-2}+\dfrac{2a^{-3/(2\omega)}}{3F}\Big[3(\rho(a)+V(a))+\omega aV^{'}(a)\Big],
\end{align}
\\and energy density is found as
\begin{align}
\rho(a)=\dfrac{3F^{2}k}{2\omega}a^{3/\omega-2}-V(a)-\dfrac{\omega aV^{'}(a)}{3}.
\end{align}
\\Examination of this singularity with $V(a)=\dfrac{V_{n}}{a^{n}}$ gives us
\begin{gather}
H(a)=\sqrt{\dfrac{\omega}{3+2\omega}}\sqrt{\dfrac{2k}{a^{2}}+\dfrac{8n\omega^{2}V_{n}}{9F^{2}a^{3/\omega+n}}},\\
\rho(a)=\dfrac{3F^{2}k}{2\omega}a^{3/\omega-2}+\dfrac{(n\omega-3)V_{n}}{3a^{n}}.
\end{gather}
\\Thus we have the energy density in usual form.\\
\indent Second singularity has been already seen in section 4.3.2. It implies $\phi(a)=Fa^{(3\pm\sqrt{9+6\omega})/(2\omega)}$.
\\Then the $\rho$ equation has the following form
\begin{align}
\dfrac{3F^{2}k}{4\omega}a^{(3-2\omega\pm \sqrt{9+6\omega})/\omega}-V(a)=\rho(a).
\end{align}
\\After writing $H(a)=\sqrt{\gamma (a)}$, $\phi$ equation becomes
\begin{align}
\gamma^{'}(a)+P(a)\gamma (a)=Q(a).
\end{align}
\\For $\phi(a)=a^{(3+\sqrt{9+6\omega})/(2\omega)}$,
\begin{gather}
P(a)=\dfrac{2(3+3\omega+\sqrt{9+6\omega})}{\omega a},\\ \nonumber \\
Q(a)=\dfrac{a^{-(3+3\omega+\sqrt{9+6\omega})/\omega}}{3\sqrt{3+2\omega}F^{2}}\Big[6\sqrt{3}F^{2}ka^{(3+\sqrt{9+6\omega})/\omega}+4\omega(\sqrt{3}-\sqrt{3+2\omega})a^{3}V^{'}(a)\Big].
\end{gather}
\\Then
\begin{gather}
\gamma(a)=\dfrac{1}{a^{2(3+3\omega+\sqrt{9+6\omega})/\omega}}\Big\{b\int_{a_{in}}^{a}a^{'(3+3\omega+\sqrt{9+6\omega})/\omega}\Big[6\sqrt{3}F^{2}ka^{'(3+\sqrt{9+6\omega})/\omega}\nonumber \\+4\omega(\sqrt{3}-\sqrt{3+2\omega})a^{'3}V^{'}(a^{'})\Big]da^{'}+\tilde{\gamma}(a_{in})\Big\},\\
b=\dfrac{1}{3\sqrt{3+2\omega}F^{2}}, \hspace{25pt}\tilde{\gamma}(a_{in})=a_{in}^{2(3+3\omega+\sqrt{9+6\omega})/\omega}\gamma(a_{in}). 
\end{gather}
\\For $\phi(a)=a^{(3-\sqrt{9+6\omega})/(2\omega)}$,
\begin{gather}
P(a)=\dfrac{2(3+3\omega-\sqrt{9+6\omega})}{\omega a},\\ \nonumber \\
Q(a)=\dfrac{a^{-3(1+\omega)/\omega}}{3F^{2}(3+2\omega-\sqrt{9+6\omega})}\Big[6F^{2}k(3-\sqrt{9+6\omega})a^{3/\omega}-8\omega^{2}a^{(3\omega+\sqrt{9+6\omega})/\omega}V^{'}(a)\Big].
\end{gather}
\\Then
\begin{gather}
\gamma(a)=\dfrac{1}{a^{2(3+3\omega-\sqrt{9+6\omega})/\omega}}\Big\{b\int_{a_{in}}^{a}a^{'(3+3\omega-2\sqrt{9+6\omega})/\omega}\Big[6F^{2}k(3-\sqrt{9+6\omega})a^{'3/\omega}\nonumber \\+8\omega^{2}a^{'(3\omega+\sqrt{9+6\omega})/\omega}V^{'}(a^{'})\Big]da^{'}+\tilde{\gamma}(a_{in})\Big\},\\
b=\dfrac{1}{3(3+2\omega-\sqrt{9+6\omega})F^{2}}, \hspace{25pt}\tilde{\gamma}(a_{in})=a_{in}^{2(3+3\omega-\sqrt{9+6\omega})/\omega}\gamma(a_{in}). 
\end{gather}
\indent At the next step we examine nature of the Hubble function. First, we take 
\begin{align}
\phi(a)=Fa^{(3+\sqrt{9+6\omega})/(2\omega)}, \hspace{20pt} V(a)=\dfrac{V_{n}}{a^{n}}.
\end{align}
\\The Hubble function and the energy density are found as
\begin{gather}
H(a)=\Big\{\dfrac{\sqrt{3}k(\sqrt{3+2\omega}-\sqrt{3})}{2(3+2\omega)a^{2}}+\dfrac{8n\omega^{3}V_{n}a^{-(3+n\omega+\sqrt{9+6\omega})/\omega}}{3(3+2\omega+\sqrt{9+6\omega})\Big[3+(6-n)\omega+\sqrt{9+6\omega}\Big]F^{2}}\nonumber \\ +\tilde{\gamma}(a_{in})a^{-2(3+3\omega+\sqrt{9+6\omega})/\omega}\Big\}^{1/2},\\
\tilde{\gamma}(a_{in})=\Big\{\gamma (a_{in})-\dfrac{\sqrt{3}k(\sqrt{3+2\omega}-\sqrt{3})}{2(3+2\omega)a^{2}_{in}} \nonumber \\-\dfrac{8n\omega^{3}V_{n}a^{-(3+n\omega+\sqrt{9+6\omega})/\omega}_{in}}{3(3+2\omega+\sqrt{9+6\omega})\Big[3+(6-n)\omega+\sqrt{9+6\omega}\Big]F^{2}}\Big\}a^{2(3+3\omega+\sqrt{9+6\omega})/\omega}_{in}, \\
\rho(a)=\dfrac{3F^{2}k}{4\omega a^{(2\omega-3-\sqrt{9+6\omega})/\omega}}-\dfrac{V_{n}}{a^{n}}.
\end{gather}
\\Although the Hubble function is real for $0< n< 6$ and $\omega\gg 1$, the energy density always has a negative component.\\
\indent Then we take
\begin{align}
\phi(a)=Fa^{(3-\sqrt{9+6\omega})/(2\omega)}, \hspace{20pt}  V(a)=\dfrac{V_{n}}{a^{n}}.
\end{align}
\\The Hubble function and the energy density are found as
\begin{gather}
H(a)=\Big\{-\dfrac{\sqrt{3}k(\sqrt{3+2\omega}+\sqrt{3})}{2(3+2\omega)a^{2}}+\dfrac{8n\omega^{3}V_{n}a^{-(3+n\omega-\sqrt{9+6\omega})/\omega}}{3(3+2\omega-\sqrt{9+6\omega})\Big[3+(6-n)\omega-\sqrt{9+6\omega}\Big]F^{2}}\nonumber \\ +\tilde{\gamma}(a_{in})a^{-2(3+3\omega-\sqrt{9+6\omega})/\omega}\Big\}^{1/2},\\
\tilde{\gamma}(a_{in})=\Big\{\gamma(a_{in})+\dfrac{\sqrt{3}k(\sqrt{3+2\omega}+\sqrt{3})}{2(3+2\omega)a^{2}_{in}}\nonumber \\-\dfrac{8n\omega^{3}V_{n}a^{-(3+n\omega-\sqrt{9+6\omega})/\omega}_{in}}{3(3+2\omega-\sqrt{9+6\omega})\Big[3+(6-n)\omega-\sqrt{9+6\omega}\Big]F^{2}}\Big\}a^{2(3+3\omega-\sqrt{9+6\omega})/\omega}_{in},\\
\rho(a)=\dfrac{3F^{2}k}{4\omega a^{(2\omega-3+\sqrt{9+6\omega})/\omega}}-\dfrac{V_{n}}{a^{n}}.
\end{gather}
\\Although the Hubble function is real for $0< n< 6$ and $\omega\gg 1$, the energy density always has a negative component.\\
\section{Solution for given $H(a)$ and $\phi(a)$}
\indent The solution for specific form of the Hubble function can be crucial. To construct the solution one can follow two different paths:
\begin{itemize}
\item Method 1) First, find $V(a)$ from the constraint equation then insert it into the $\phi$ equation and solve it for $\rho(a)$.
\item Method 2) First, solve the $\phi$ equation for the $V(a)$ then insert it into the constraint equation and obtain $\rho(a)$.
\end{itemize}
\subsection{Method1}
\indent We find $V(a)$ from the $\rho$ equation
\begin{align}
V(a)=-\rho(a)+\dfrac{3k\phi^{2}(a)}{4\omega a^{2}}+\Big[3\phi^{2}(a)+6a\phi(a)\phi^{'}(a)-2\omega a^{2}\phi^{'2}(a)\Big]\dfrac{H^{2}(a)}{4\omega}.
\end{align}
\\Then the $\phi$ equation turns into the following form
\begin{gather}
\rho^{'}(a)=Q(a),\\
Q(a)=\dfrac{-3\phi^{2}(k-a^{3}HH^{'})+a^{3}[3(1+2\omega)aH^{2}\phi^{'2}]+3a^{4}H\phi(H^{'}\phi^{'}+H\phi^{''})}{2\omega a^{3}}.
\end{gather}
\\Thus 
\begin{gather}
\rho(a)=\int_{a_{in}}^{a}Q(a^{'})da^{'}+\rho(a_{in}), \\
V(a)=-\Big[\int_{a_{in}}^{a}Q(a^{'})da^{'}+\rho(a_{in})\Big]+\dfrac{3k\phi^{2}(a)}{4\omega a^{2}}+\Big[3\phi^{2}(a)+6a\phi(a)\phi^{'}(a)-2\omega a^{2}\phi^{'2}(a)\Big]\dfrac{H^{2}(a)}{4\omega}.
\end{gather}
\subsection{Method2}
\indent We solve the $\phi$ equation for the potential,
\begin{align}
V(a)=-\int_{a_{in}}^{a}\phi^{'}\Big\{-3\phi\Big[k+a^{'2}H(2H+a^{'}H^{'})\Big]+2\omega a^{'3}H\Big[(4H+a^{'}H^{'})\phi^{'}+a^{'}H\phi^{''}\Big]\Big\}da^{'}+V(a_{in}).
\end{align}
\\We obtain energy density by substituting $V(a)$ into the $\rho$ equation
\begin{gather}
\rho(a)=\int_{a_{in}}^{a}\phi^{'}\Big\{-3\phi\Big[k+a^{'2}H(2H+a^{'}H^{'})\Big]+2\omega a^{'3}H\Big[(4H+a^{'}H^{'})\phi^{'}+a^{'}H\phi^{''}\Big]\Big\}da^{'}-V(a_{in})\nonumber \\+\dfrac{3k\phi^{2}(a)}{4\omega a^{2}}+\Big[3\phi^{2}(a)+6a\phi(a)\phi^{'}(a)-2\omega a^{2}\phi^{'2}(a)\Big]\dfrac{H^{2}(a)}{4\omega}.
\end{gather}
\section{Single Component Universe}
\indent We investigate the universe with given energy density and the scalar field which are in the following form
\begin{align}
\rho(a)=\dfrac{\rho_{n}}{a^{n}},\hspace{30pt} \phi(a)=\dfrac{F}{a^{d}}.
\end{align}
\\We have obtained the Hubble function and the potential function by applying the procedure explained in section 4.1, we obtain the following formulas
\begin{gather}
H(a)=\sqrt{\dfrac{k}{\mathcal{B}a^{2}}-\dfrac{4n\rho_{n}\omega a^{2d-n}}{3\mathcal{C}F^{2}}+\tilde{\gamma}(a_{in})a^{\mu}},\\
V(a)=\dfrac{d^{2}(3+2\omega)F^{2}k}{2\mathcal{B}\omega a^{2+2d}}+\dfrac{\mathcal{A}\rho_{n}a^{-n}}{3\mathcal{C}}+\dfrac{\mathcal{F}\tilde{\gamma}(a_{in})F^{2}a^{-2d+\mu}}{4\omega},\\
\tilde{\gamma}(a_{in})=a_{in}^{-\mu}H^{2}(a_{in})-\dfrac{ka^{-2-\mu}_{in}}{\mathcal{B}}+\dfrac{4n\rho_{n}\omega a^{2d-n-\mu}_{in}}{3\mathcal{C}F^{2}},\\
\mu=\dfrac{2d[1+2d(1+\omega)]}{-1+d},\\
\mathcal{A}=d\{3(-4+n)+2d[-3+(-6+n)\omega]\}, \\
\mathcal{B}=[-1+2d(1+d+d\omega)],\\
\mathcal{C}=\{-n+d[4+n+d(2+4\omega)]\},\\
\mathcal{F}=[-3+2d(3+d\omega)],\\
p(a)=\dfrac{(n-3)\rho_{n}}{3a^{n}}.
\end{gather}
\\We should choose $d$ such that $\mathcal{B}<0$ and $\mathcal{C}<0$. Then we have real Hubble function. In addition if we also satisfy $\mu=0$ at the same time this will have  important physical implications which will be discussed soon. Thus
\begin{gather}
\mu=0 \hspace{20pt}\Rightarrow \hspace{20pt} d=-\dfrac{1}{2(1+\omega)},\\
\mathcal{B}=-1-\dfrac{1}{2(1+\omega)},\hspace{20pt}\mathcal{C}=-\dfrac{(3+2\omega)(1+n+n\omega)}{2(1+\omega)^{2}}.
\end{gather}
\\Then the Hubble function and potential are simplified to the following forms
\begin{gather}
H(a)=\sqrt{-\dfrac{2(1+\omega)k}{(3+2\omega)a^{2}}+\dfrac{8n(1+\omega)^{2}\omega\rho_{n}a^{-1/(1+\omega)-n}}{3(3+2\omega)[1+n(1+\omega)]F^{2}}+\tilde{\gamma}(ain)},\\
V(a)=\dfrac{(-3+n)\rho_{n}a^{-n}}{3(1+n(1+\omega))}-\dfrac{kF^{2}a^{1/(1+\omega)-2}}{4\omega (1+\omega)}+\dfrac{(3+2\omega)(4+3\omega)F^{2}}{8\omega(1+\omega)^{2}}\tilde{\gamma}(a_{in})a^{1/(1+\omega)},\\
\tilde{\gamma}(a_{in})=H^{2}(a_{in})+\dfrac{2(1+\omega)k}{(3+2\omega)a^{2}_{in}}-\dfrac{8n(1+\omega)^{2}\omega\rho_{n}a^{-1/(1+\omega)-n}_{in}}{3(3+2\omega)[1+n(1+\omega)]F^{2}}.
\end{gather}
\indent The deceleration parameter is found by applying the chain rule
\begin{gather}
q(t)=\dfrac{d}{dt}(\dfrac{1}{H(t)})-1,\\
q(a)=\dfrac{d}{da}(\dfrac{1}{H(a)})aH(a)-1,\\
q(a)=-1+\dfrac{2(1+\omega)[1+n(1+\omega)][-3F^{2}ka^{1/(1+\omega)+n}+2n\omega\rho_{n}a^{2}]}{8n\omega(1+\omega)^{2}\rho_{n}a^{2}+3[1+n(1+\omega)]F^{2}[-2k(1+\omega)+(3+2\omega)\tilde{\gamma}(a_{in})a^{2}]a^{1/(1+\omega)+n}}.
\end{gather}
\\When finding $q(a_{in})$ one should plug $\tilde{\gamma}(a_{in})$ which is given by (127)
\begin{align}
q(a_{in})=-1+\dfrac{2(1+\omega)[-3F^{2}ka^{1/(1+\omega)+n}_{in}+2n\omega\rho_{n}a^{2}_{in}]}{3(3+2\omega)F^{2}H^{2}(a_{in})a^{2}_{in}}.
\end{align}
\\Since $a_{in}\ll 1$, $q(a_{in})$ is simplified further
\begin{gather}
\lim_{a\rightarrow a_{in}}q(a)=
\begin{cases}
-1+\dfrac{4(1+\omega)n\omega\rho_{n}}{3(3+2\omega)F^{2}H^{2}(a_{in})},\hspace{20pt} if \hspace{20pt} n+\dfrac{1}{1+\omega}>2, \nonumber\\ \nonumber \\
-1+\dfrac{2(1+\omega)[-3F^{2}k+2n\omega\rho_{n}]}{3(3+2\omega)F^{2}H^{2}(a_{in})}
 \hspace{20pt} if \hspace{20pt} n+\dfrac{1}{1+\omega}=2, \nonumber \\ \nonumber \\
-1-\dfrac{2(1+\omega)k}{(3+2\omega)H^{2}(a_{in})a^{-1/(1+\omega)-n+2}_{in}},\hspace{20pt} if \hspace{20pt} 0<n+\dfrac{1}{1+\omega}<2.
\end{cases}
\end{gather}
\\When we set today values of the scale factor $a=1$, $\rho_{n}$ and $F$ become today's values of the energy density and the scalar field respectively. Thus dynamics of the early universe depends on not only initial value of the Hubble function and the scale factor but also today's values of the energy density and scalar field. In addition behaviour of the universe also depends on the curvature parameter $k$  for $0<n+\dfrac{1}{1+\omega}\leq 2$. 
\\The fate of the universe can be described by the following number
\begin{align}
\lim_{a\rightarrow \infty}q(a)=-1.
\end{align}
\indent Now, we formulate $a(t)$ for a spatially flat universe as
\begin{gather}
t=\int_{a_{in}}^{a}\dfrac{da^{'}}{a^{'}H(a^{'})},\\
H(a)=\sqrt{\delta a^{-1/(1+\omega)-n}+\alpha}, \hspace{20pt}\delta=\dfrac{8n(1+\omega)^{2}\omega\rho_{n}}{3(3+2\omega)[1+n(1+\omega)]F^{2}},\hspace{20pt}\alpha=\tilde{\gamma}(a_{in}).
\end{gather}
\\Thus
\begin{align}
t=\dfrac{1}{\kappa \sqrt{\alpha}}ln\Big[\dfrac{a^{\kappa}+\sqrt{a^{2\kappa}+\dfrac{\delta}{\alpha}}}{a^{\kappa}_{in}+\sqrt{a^{2\kappa}_{in}+\dfrac{\delta}{\alpha}}}\Big],\hspace{25pt} \kappa =\dfrac{1}{2}\Big[\dfrac{1+n(1+\omega)}{1+\omega}\Big].
\end{align}
\\Hence
\begin{align}
a(t)=\Big[\dfrac{(a^{\kappa}_{in}+\sqrt{a^{2\kappa}_{in}+\dfrac{\delta}{\alpha}})e^{\kappa\sqrt{\alpha}t}+(a^{\kappa}_{in}-\sqrt{a^{2\kappa}_{in}+\dfrac{\delta}{\alpha}})e^{-\kappa\sqrt{\alpha}t}}{2}\Big]^{1/\kappa},
\end{align}
\\where $\delta$, $\alpha$ and $\kappa$ are given in (134-135). This formula indicates a very significant interpretation.  When there is no constant energy density, the universe still expands exponentially because of nonzero initial conditions. We will investigate this phenomena in depth by adding constant term to the energy density in sections 8 and 9\\
\indent As we said at the beginning of section 2 we have obtained perfect fluid equation of state
\begin{align}
\nu=\dfrac{p(a)}{\rho(a)}=\dfrac{(n-3)}{3}.
\end{align}
\indent Potential formula as a function of the scalar field are written as
\begin{align}
V(\phi)=\dfrac{(-3+n)\rho_{n}}{3(1+n(1+\omega))}(\dfrac{\phi}{F})^{-2n(1+\omega)}-\dfrac{kF^{2}}{4\omega(1+\omega)}(\dfrac{\phi}{F})^{-2-4\omega}+\dfrac{(3+2\omega)(4+3\omega)}{8\omega(1+\omega)^{2}}\tilde{\gamma}(a_{in})\phi^{2}.
\end{align}
\\Most probably $\phi^{2}$ term in the potential is responsible for late time accelerated expansion. This formula is simpler than the corresponding formula $V(\phi)$ in our previous study \cite{ildes2022analytic}.\\
\section{Early Universe, Dark Energy and Radiation}
\indent In this section the scalar field and the energy density is taken as
\begin{align}
\phi(a)=\dfrac{F}{a^{d}}, \hspace{20pt}d=-\dfrac{1}{2(1+\omega)}, \hspace{20pt} \rho(a)=\dfrac{\rho_{r}}{a^{4}}+\Lambda.
\end{align}
\\To see whether $\Lambda$ contributes to $H(a)$ we have performed calculations in all four different methods. Results are found in the following form
\begin{gather}
H(a)=\sqrt{-\dfrac{2(1+\omega)k}{(3+2\omega)a^{2}}+\dfrac{32(1+\omega)^{2}\omega\rho_{r}a^{-1/(1+\omega)-4}}{3(3+2\omega)(5+4\omega)F^{2}}+\alpha}, \\
V(a)=-\dfrac{F^{2}ka^{1/(1+\omega)-2}}{4\omega(1+\omega)}+\beta a^{1/(1+\omega)}+\dfrac{\rho_{r}}{3(5+4\omega)a^{4}}-\Lambda.
\end{gather}
\indent In method-1 we find
\begin{gather}
\alpha_{1}=H^{2}(a_{in})+\dfrac{2(1+\omega)k}{(3+2\omega)a_{in}^{2}}-\dfrac{32(1+\omega)^{2}\omega\rho_{r}a^{-1/(1+\omega)-4}_{in}}{3(3+2\omega)(5+4\omega)F^{2}},\\
\beta_{1}=\dfrac{(3+2\omega)(4+3\omega)F^{2}}{8\omega(1+\omega)^{2}}\alpha_{1}.
\end{gather}
\indent In method-2 we find
\begin{gather}
\alpha_{2}=\alpha_{1}, \hspace{20pt} \beta_{2}=\beta_{1}.
\end{gather}
\\In this method to be able to satisfy the constraint equation with final formulas of $V(a)$ and $H(a)$ initial conditions must satisfy the following equation
\begin{align}
V(a_{in})=-\dfrac{F^{2}ka^{1/(1+\omega)-2}_{in}}{4\omega(1+\omega)}+\beta_{2} a^{1/(1+\omega)}_{in}+\dfrac{\rho_{r}}{3(5+4\omega)a^{4}_{in}}-\Lambda,
\end{align}
\\which is consistent with the formula of $V(a)$.\\
\indent In method-3 we find
\begin{gather}
\alpha_{3}=\dfrac{8\omega(1+\omega)^{2}\tilde{V}(a_{in})}{(3+2\omega)(4+3\omega)F^{2}},\\
\tilde{V}(a_{in})=V(a_{in})a^{-1/(1+\omega)}_{in}-\dfrac{\rho_{r}a^{-1/(1+\omega)-4}_{in}}{3(5+4\omega)}+\dfrac{F^{2}k}{4\omega(1+\omega)a^{2}_{in}}+\Lambda a^{-1/(1+\omega)}_{in},  \\
\beta_{3}=\dfrac{(3+2\omega)(4+3\omega)F^{2}}{8\omega(1+\omega)^{2}}\alpha_{3}.
\end{gather}
\\Although at first sight it seems that $\alpha_{1}\neq\alpha_{3}$, one can show that they are equal to each other by substituting $a_{in}$ in $H(a)$ and solving the equation for $V(a_{in})$.\\
\indent In method-4 we find
\begin{gather}
\alpha_{4}=\alpha_{3}, \hspace{20pt} \beta_{4}=\beta_{3}.
\end{gather}
\\By the same reasoning in method-2, constraint equation implies the following condition on initial values
\begin{align}
a^{1/(1+\omega)+3}_{in}H^{2}(a_{in})-\dfrac{32\omega(1+\omega)^{2}\rho_{r}}{3(3+2\omega)(5+4\omega)F^{2}a_{in}}-\dfrac{8\omega(1+\omega)^{2}\tilde{V}(a_{in})a^{1/(1+\omega)+3}_{in}}{(3+2\omega)(4+3\omega)F^{2}}+\dfrac{2(1+\omega)ka^{1/(1+\omega)+1}_{in}}{3+2\omega}=0.
\end{align}
\\$V(a_{in})$ which is obtained by solving this equation is consistent with the formula of $V(a)$.\\
\indent The deceleration parameter is found as
\begin{gather}
q(a)=-1+\dfrac{2(1+\omega)(5+4\omega)[-3F^{2}ka^{1/(1+\omega)+2}+8\omega\rho_{r}]}{32\omega(1+\omega)^{2}\rho_{r}+3(5+4\omega)F^{2}[-2k(1+\omega)+(3+2\omega)\alpha_{1}a^{2}]a^{1/(1+\omega)+2}},\\
\alpha_{1}=H^{2}(a_{in})+\dfrac{2(1+\omega)k}{(3+2\omega)a^{2}_{in}}-\dfrac{32(1+\omega)^{2}\omega\rho_{r}a^{-1/(1+\omega)-4}_{in}}{3(3+2\omega)(5+4\omega)F^{2}}.
\end{gather}
\\Thus initial value of the deceleration parameter is found
\begin{align}
q(a_{in})=-1+\dfrac{2(1+\omega)[-3F^{2}ka^{1/(1+\omega)+2}_{in}+8\omega\rho_{r}]}{3(3+2\omega)F^{2}H^{2}(a_{in})}.
\end{align}
\indent The deceleration parameter which is given in (151) is same as the deceleration parameter found in single component universe without constant term, since $\alpha_{1}$ given in (152) is equal to $\tilde{\gamma}(a_{in})$ given in (127) with $n=4$.\\
\indent The scale factor as a function of time is written as
\begin{gather}
a(t)=\Big[\dfrac{(a^{\kappa}_{in}+\sqrt{a^{2\kappa}_{in}+\dfrac{\delta}{\alpha}})e^{\kappa\sqrt{\alpha}t}+(a^{\kappa}_{in}-\sqrt{a^{2\kappa}_{in}+\dfrac{\delta}{\alpha}})e^{-\kappa\sqrt{\alpha}t}}{2}\Big]^{1/\kappa},\\
\kappa=\dfrac{1}{2}\Big(\dfrac{5+4\omega}{1+\omega}\Big), \hspace{20pt} \delta=\dfrac{32(1+\omega)^{2}\omega\rho_{r}}{3(3+2\omega)(5+4\omega)F^{2}},
\end{gather}
\\where $\alpha=\alpha_{1}$ and $\alpha_{1}$ is given by (142).\\
\indent Formula of the potential as function of the scalar field is found as
\begin{align}
V(\phi)=-\dfrac{F^{2}k}{4\omega(1+\omega)}(\dfrac{\phi}{F})^{-2-4\omega}+\beta (\dfrac{\phi}{F})^{2} +\dfrac{\rho_{r}}{3(5+4\omega)}(\dfrac{\phi}{F})^{-8(1+4\omega)}-\Lambda,
\end{align}
\\where $\beta=\beta _{1}$ which is given by (143). Pressure is found as in expected form
\begin{align}
p(a)=\dfrac{\rho_{r}}{3a^{4}}-\Lambda.
\end{align}
\section{Late Time expansion of the universe}
\subsection{Dark Energy(!) dominated universe}
\indent We study in spatially flat universe where
\begin{align}
\phi(a)=\dfrac{F}{a^{d}}, \hspace{20pt} d=-\frac{1}{2(1+\omega)},\hspace{20pt}\rho(a)=\dfrac{\rho_{m}}{a^{3}}+\Lambda.
\end{align}
\\Then related function is formulated as
\begin{gather}
H(a)=\sqrt{\dfrac{8\omega(1+\omega)^{2}\rho_{m}}{(3+2\omega)(4+3\omega)F^{2}a^{1/(1+\omega)+3}}+\tilde{\gamma}(a_{in})},\\
V(a)=\dfrac{(3+2\omega)(4+3\omega)}{8\omega(1+\omega)^{2}}F^{2}\tilde{\gamma}(a_{in})a^{1/(1+\omega)}-\Lambda,\\
q(a)=-1+\dfrac{4\omega(1+\omega)(4+3\omega)\rho_{m}}{8\omega(1+\omega)^{2}\rho_{m}+(3+2\omega)(4+3\omega)F^{2}\tilde{\gamma}(a_{in})a^{1/(1+\omega)+3}},\\
\tilde{\gamma}(a_{in})=H^{2}(a_{in})-\dfrac{8(1+\omega)^{2}\omega\rho_{m}a^{-1/(1+\omega)-3}_{in}}{(3+2\omega)(4+3\omega)F^{2}},\\
p(a)=-\Lambda.
\end{gather}
\indent The scale factor as function of time is found as
\begin{gather}
a(t)=\Big[\dfrac{(a^{\kappa}_{in}+\sqrt{a^{2\kappa}_{in}+\dfrac{\delta}{\alpha}})e^{\kappa\sqrt{\alpha}t}+(a^{\kappa}_{in}-\sqrt{a^{2\kappa}_{in}+\dfrac{\delta}{\alpha}})e^{-\kappa\sqrt{\alpha}t}}{2}\Big]^{1/\kappa},\\
\kappa=\dfrac{1}{2}\Big(\dfrac{4+3\omega}{1+\omega}\Big), \hspace{20pt} \delta=\dfrac{8(1+\omega)^{2}\omega\rho_{m}}{(3+2\omega)(4+3\omega)F^{2}}.
\end{gather}
\\where $\alpha=\tilde{\gamma}(a_{in})$ and $\tilde{\gamma}(a_{in})$ is given by (162).\\
\indent In Brans-Dicke theory effective gravitational constant is defined as $G_{eff}=\dfrac{\omega}{2\pi \phi^{2}}$. Thus it's present value becomes $G_{0}=\dfrac{\omega}{2\pi F^{2}}$ where $F=\phi(t_{0})$. In addition when we take observational value of Brans-Dicke parameter $\omega\gg 1$, the Hubble function becomes
\begin{align}
H(a)=\sqrt{\dfrac{8\pi G_{0}\rho_{m}}{3a^{3}}+\tilde{\gamma}(a_{in})}.
\end{align}
\\By using following definitions of cosmological parameters 
\begin{align}
\Omega_{m,0}=\dfrac{8\pi G}{3H^{2}_{0}}\rho_{m}, \hspace{20pt}\Omega_{\tilde{\gamma},0}=\dfrac{8\pi G}{3H^{2}_{0}}\rho_{\tilde{\gamma}},\hspace{20pt} \rho_{\tilde{\gamma}}=\dfrac{3\tilde{\gamma}(a_{in})}{8\pi G},
\end{align}
\\the Hubble function is written as
\begin{align}
H(a)=H_{0}\sqrt{\dfrac{\Omega_{m,0}}{a^{3}}+\Omega_{\tilde{\gamma},0}}.
\end{align}
\\When one replace $\Omega_{\tilde{\gamma},0}$ with $\Omega_{\Lambda,0}$ in (166) we will have the Hubble function for the late epoch of the universe in standard cosmology where dark energy dominates.\\
\indent We have already compared this model with observation of type Ia supernovae data in \cite{ildes2022analytic}. In our previous study we have obtained $H_{0}=71.80\pm 0.22$, $\Omega_{\Lambda,0}=0.715\pm 0.012$, $\Omega_{m}=0.285\pm 0.012$ and $\chi^{2}/\nu=0.990$ for absolute magnitude $M=-19.30$. Present value of deceleration parameter has been found as $q_{0}=-0.572$ with this cosmological parameters. \\
\indent On the other hand potential can be rewritten as
\begin{align}
V(\phi)=\dfrac{(3+2\omega)(4+3\omega)}{8\omega(1+\omega)^{2}}\tilde{\gamma}(a_{in})\phi^{2}-\Lambda.
\end{align}
\\Most probably $\phi^{2}$ potential is responsible from the accelerated expansion.\\
\subsection{Domain wall dominated universe}
\indent In this section we study in spatially flat universe with scalar field and the energy density is given as
\begin{align}
\phi(a)=\dfrac{F}{a^{d}}, \hspace{20pt}d=-\dfrac{1}{2(1+\omega)}, \hspace{20pt} \rho(a)=\dfrac{\rho_{m}}{a^{3}}+\dfrac{\rho_{w}}{a}.
\end{align}
\\The Hubble function, potential, deceleration parameter and the pressure is formulated as
\begin{gather}
H(a)=\sqrt{\dfrac{8\omega(1+\omega)^{2}}{3(3+2\omega)F^{2}}\Big[\dfrac{3\rho_{m}}{(4+3\omega)a^{3}}+\dfrac{\rho_{\omega}}{(2+\omega)a}\Big]a^{-1/(1+\omega)}+\tilde{\gamma}(a_{in})},\\
V(a)=\dfrac{(3+2\omega)(4+3\omega)}{8\omega(1+\omega)^{2}}\tilde{\gamma}(a_{in})F^{2}a^{1/(1+\omega)}-\dfrac{2\rho_{w}}{3(2+\omega)a},\\
q(a)=-1+\dfrac{4\omega(1+\omega)(2+\omega)(4+3\omega)(3\rho_{m}+\rho_{w}a^{2})}{3(2+\omega)(3+2\omega)(4+3\omega)\tilde{\gamma}(a_{in})F^{2}a^{1/(1+\omega)+3}+8\omega(1+\omega)^{2}[3(2+\omega)\rho_{m}+(4+3\omega)\rho_{w}a^{2}]},\\
\tilde{\gamma}(a_{in})=H^{2}(a_{in})-\dfrac{8\omega(1+\omega)^{2}}{3(3+2\omega)F^{2}}\Big[\dfrac{3\rho_{m}}{(4+3\omega)a^{3}_{in}}+\dfrac{\rho_{w}}{(2+\omega)a_{in}}\Big]a^{-1/(1+\omega)}_{in},\\
p(a)=-\dfrac{2\rho_{w}}{3a}.
\end{gather}
\indent Now we take $\tilde{\gamma}(a_{in})=0$, to investigate whether cosmic domain walls cause accelerated expansion of the universe or not. We again apply the fundamental idea of the Brans-Dicke theory as we did in section 8. Thus we take $G_{0}=\dfrac{\omega}{2\pi F^{2}}$ and we take $\omega\gg 1$. Then the Hubble function is simplified to
\begin{align}
H(a)=\sqrt{\dfrac{8\pi G_{0}}{3}(\dfrac{\rho_{m}}{a^{3}}+\dfrac{\rho_{w}}{a})}.
\end{align}
\indent The relation between the scale factor and time is found as
\begin{gather}
t=\int_{a_{in}}^{a}\dfrac{da^{'}}{a^{'}H(a^{'})}, \nonumber \\
t=\int_{a_{in}}^{a}\dfrac{da^{'}}{a^{'}\sqrt{\dfrac{8\pi G}{3}(\dfrac{\rho_{m}}{a^{3}}+\dfrac{\rho_{w}}{a})}}, \\
t=\sqrt{\dfrac{3}{2\pi G\rho_{w}}}\Big[\sqrt{a^{'}}{}_{2}F_{1}(\dfrac{1}{2},-\dfrac{1}{4};\dfrac{3}{4};-\dfrac{\rho_{m}}{\rho_{w} a^{'2}})\Big]_{a_{in}}^{a}.
\end{gather}
\\${}_{2}F_{1}$ is the hypergeometric function 
\begin{align}
{}_{2}F_{1}(1/2,b;b+1;u)=\sum_{n=0}^{\infty}\dfrac{(1/2)_{n}(b)_{n}}{(b+1)_{n}}\dfrac{u^{n}}{n!},
\end{align}
\\where $(b)_{n}$ is the Pochhammer symbol which is defined as
\begin{align}
(b)_{n}=
\begin{cases}
1 \hspace{20pt} \text{if} \hspace{20pt} n=0, \\
b(b+1)(b+2)...(b+n-1),\hspace{20pt} \text{if} \hspace{20pt} n>0. 
\end{cases}
\end{align}
\indent Type Ia supernovaes are known as standard candles since their measurements makes comparison of theory and observations is possible. This is done by the following relation
\begin{align}
m=5log_{10}(\dfrac{d_{L}}{1Mpc})+25+M,
\end{align}
\\where $m$ and $M$ are the apparent and the absolute magnitudes respectively. Then the distance modulus is defined as $\mu=m-M$.\\
\indent  Now we borrow one of our plots from our previous work \cite{ildes2022analytic}. In our previous study we have already performed curve fitting with type Ia supernovae data. We have found $H_{0}=71.03\pm 0.20$, $\Omega_{w,0}=0.889\pm 0.015$, $\Omega_{m,0}=0.111\pm 0.015$ and $\chi^{2}/\nu=1.008$ for $M=-19.30$ in domain wall dominated universe. We obtain $q_{0}=-0.389$ with these values of cosmological density parameters. On the other hand for dark energy dominated universe we have found $H_{0}=71.80\pm 0.22$, $\Omega_{\Lambda,0}=0.715\pm 0.012$, $\Omega_{m,0}=0.285\pm 0.012$ and $\chi^{2}/\nu=0.990$ for $M=-19.30$. We obtain $q_{0}=-0.572$. All these results are shown in Figure 1.\\
\begin{figure}[htb]
        \includegraphics[width=14cm, height=8cm]{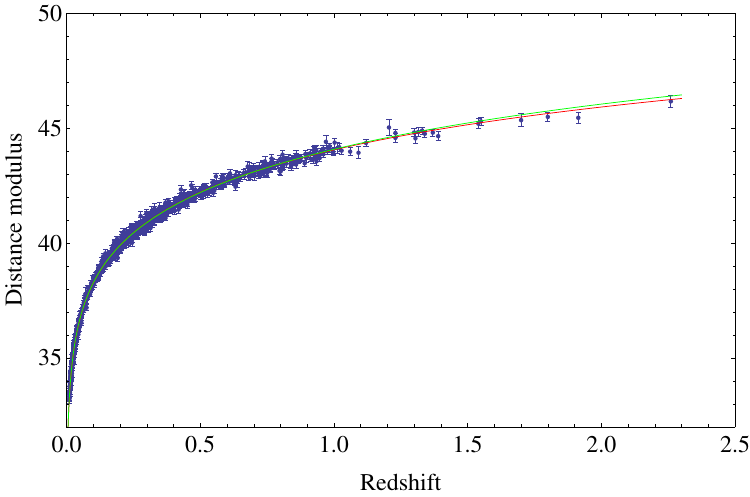}
     \caption{Figure I:Distance modulus vs redshift plot for $M=-19.30$. Dots represent observations of Pantheon data while green line represents domain wall dominated universe and red line represents dark energy dominated universe}
  \end{figure}
\\In addition
\begin{align}
\lim_{a\Rightarrow \infty}q(a)=-0.5.
\end{align}
\indent $ V(\phi)$ is formulated as
\begin{align}
V(\phi)=-\dfrac{2\rho_{w}}{3(2+\omega)}(\dfrac{\phi}{F})^{-2(1+\omega)}.
\end{align}
\\Although power of $\phi$ in the potential is different from $2$ it causes accelerated expansion of the universe. \\
\section{Discussion}
\indent We have revealed three major points of cosmology in this study. These are the answers for the following questions:\\
\begin{itemize}
\item[1] Is there any corresponding energy for the cosmological constant?
\item[2] Is it possible to have accelerated universe without cosmological constant?
\item[3] Is ratio of dark mater to baryonic matter smaller than $\dfrac{0.27}{0.05}$?
\end{itemize}
\indent It is known that constant energy density contributes to the Hubble function. We have examined this phenomena in section 8 and in section 9.1. Firstly we have solved our system of equations for early universe by taking 
\begin{align}
\phi(a)=\dfrac{F}{a^{d}}, \hspace{20pt} d=-\dfrac{1}{2(1+\omega)}, \hspace{20pt} \rho(a)=\dfrac{\rho_{r}}{a^{4}}+\Lambda.
\end{align} 
\\Then $H(a)$ is found in the form of (140) where there is a constant term $\alpha$. At first sight there is no contribution of $\Lambda$ in $\alpha_{1}$ given by (142). However one can also claim that $\Lambda$ contributes in $\alpha_{3}$ by substituting (147) in (146). This dilemma can be explained as follows. One can write 
\begin{align}
V(a_{in})=\dfrac{3F^{2}ka^{1/(1+\omega)-2}_{in}}{4\omega}+\dfrac{(3+2\omega)(4+3\omega)}{8\omega(1+\omega)^{2}}F^{2}H^{2}(a_{in})a^{1/(1+\omega)}(a_{in})-\dfrac{\rho_{r}}{a^{4}_{in}}-\Lambda.
\end{align}
\\Then one can rewrite $H(a_{in})$ in terms of $V(a_{in})$ and $\Lambda$.\\
\indent We have also examined the spatially flat late universe in section 9.1 by supplying the following functions
\begin{align}
\phi(a)=\dfrac{F}{a^{d}}, \hspace{20pt} d=-\dfrac{1}{2(1+\omega)}, \hspace{20pt} \rho(a)=\dfrac{\rho_{m}}{a^{3}}+\Lambda.
\end{align}
\\The Hubble function has been found in the form given in (159) which has a constant term.  At first sight this constant term which is given by  (162) does not contain $\Lambda$. One can apply the same trick and one can write
\begin{align}
V(a_{in})=\dfrac{(3+2\omega)(4+3\omega)}{8\omega(1+\omega)^{2}}F^{2}H^{2}(a_{in})a^{1/(1+\omega)}_{in}-\dfrac{\rho_{m}}{a^{3}_{in}}-\Lambda.
\end{align}
\\Again, one can rewrite $H(a_{in})$ in terms of $V(a_{in})$ and $\Lambda$ and one can claim that $H(a)$ contains $\Lambda$.\\
\indent There are two possible interpretations;
\begin{itemize}
\item[I] Constant energy density does not contribute to the Hubble function, it only modifies value of potential at the beginning of the universe.
\item[II] Initial value of the Hubble function $H(a_{in})$, can be rewritten in terms of $V(a_{in})$ and $\Lambda$ so constant energy density contributes to the Hubble function.
\end{itemize}
\indent Both of them can be reasonable according to one's perspective.\\
\indent Although if one chooses second explanation which is in agreement with common trend, one can still ask the question in another way: \emph{Is it possible to have a constant term in the Hubble function when there is no constant energy density in the universe?} To answer this question we have investigated single component universe in section 7 by giving following functions
\begin{align}
\phi(a)=\dfrac{F}{a^{d}}, \hspace{20pt} d=-\dfrac{1}{2(1+\omega)}, \hspace{20pt} \rho(a)=\dfrac{\rho_{n}}{a^{n}}.
\end{align}
\\Then $H(a)$ is found in the form given by (125). The Hubble function has a constant term although there is no constant energy density. This constant is given in (127). Now  $V(a_{in})$ does not contain $\Lambda$. In addition, expected exponential expansion of $a(t)$ is given in (136). Thus there is one possible explanation. When there is no constant energy density, the Hubble function still has a constant term which causes exponential expansion.\\
\indent We have shown that universe has an accelerated expansion when one introduces cosmic domain walls with matter in the energy density in section 9.2. We had studied this case with Friedmann cosmology in \cite{ildes2022analytic}, and we have compared our model with the latest supernovae data. It is seen that both dark energy dominated and domain wall dominated universe have perfect fit with data. The differences between these two models most probably will be seen when bigger redshift data are available.\\
\indent While comparison of dark energy dominated universe with observation has resulted in $\Omega_{\Lambda}=0.715\pm 0.012$ and $\Omega_{m}=0.285\pm 0.012$, comparison of domain wall dominated universe with observation has resulted in $\Omega_{w}=0.889\pm 0.015$ and $\Omega_{m}=0.111\pm 0.015$. Hence a new question appears; is the ratio of dark matter to baryonic matter less than $\dfrac{0.27}{0.05}$? \\
\indent In 1980s Modified theories of Newtonian Dynamics or MOND, were proposed \cite{milgrom1983modification,milgrom1983modification2}. It mainly claims that observational aspects of galaxies can be understood without dark matter. Recent observational evidence for the external field effect in MOND which was proposed as an alternative to dark matter is presented in \cite{chae2020testing}. A new relativistic MOND theory \cite{skordis2021new} successfully reproduces cosmic microwave background power spectra.  These theories may explain non-baryonic part of domain wall dominated universe which covers six percent of the energy-matter content of the universe.\\
\section{Conclusion}
\indent In this study we have applied change of variable $a=\dot{a}H(a)$ to field equations of Brans-Dicke theory and have written all equations in terms of independent variable $"a"$. Then we have ended up with a constraint equation and a Bernoulli type differential equation which can be linearized. We have presented analytic solutions for supplied pairs of functions; ($\phi(a)$ and  $\rho(a)$) , ($\phi(a)$ and  $V(a)$), or ($\phi(a)$ and  $H(a)$).\\
\indent Investigation of single component universe has shown that one will have a constant term in the Hubble function although there is no constant energy density. Early epoch of the universe with dark energy and radiation have been studied, and exponential expansion is seen in $a(t)$. Late-time acceleration is obtained for both dark energy dominated universe and domain wall dominated universe. When Brans-Dicke parameter $\omega\gg 1$, the Hubble function reduces to $H(a)$ of the Einstein cosmology. In all cases potential is found as combination of power law potentials.\\
\indent We would like to emphasize that so-called dark energy term may be just a number without corresponding constant energy-matter density. Details of this subject has been presented in the discussion. Hopefully astronomers and cosmologists will pay more attention for searches of cosmic domain walls. 
\section*{Acknowledgement}
\indent We thank Teoman Turgut for reading first draft of this article and for fruitfull discussion. We thank Bogazici University for the financial support provided by the Scientific Research Fund (BAP), research project No 16521.

\setcounter{equation}{0}
\renewcommand{\theequation}{A\arabic{equation}}

\begin{appendices}
\section{Change of Independent Variable}
\setcounter{equation}{0}
\renewcommand{\theequation}{A\arabic{equation}}
\begin{gather}
\dot{a}=Ha \\
\frac{\dot{a}^{2}}{a^{2}}=H^{2}\\
\frac{\ddot{a}}{a}=H^{2}+H^{'}H a \\
\dot{\phi}=\phi^{'}Ha \\
\ddot{\phi}=\phi^{''}H^{2}a^{2}+\phi^{'}H^{'}H a^{2}+\phi^{'}H^{2}a \\
\end{gather}
\\where the prime denotes $\dfrac{d}{da}$.
\section{The Linear First Order Differential Equation}
\setcounter{equation}{0}
\renewcommand{\theequation}{B\arabic{equation}}
\indent The initial value problem
\begin{align}
\dfrac{dy}{dx}+P(x)y=Q(x), \hspace{20pt} y(x_{0})=y_{0},
\end{align}
has the following solution \cite{edwards2000differential}
\begin{align}
y(x)&=\dfrac{1}{\rho(x)}\Big[\int_{x_{0}}^{x}\rho(x^{'})Q(x^{'})dx^{'}+y_{0}\Big],\\
\rho(x)&=exp\Big(\int_{x_{0}}^{x} P(x^{'})dx^{'}\Big).\nonumber
\end{align}
\section{The Bernoulli Equation}
\setcounter{equation}{0}
\renewcommand{\theequation}{C\arabic{equation}}
\indent A first order differential equation which is in the following form
\begin{align}
\dfrac{dy}{dx}+P(x)y=Q(x)y^{n},
\end{align}
is known as a Bernoulli equation. It is linearized by a transformation
\begin{align}
v=y^{1-n}.
\end{align}
Thus the original nonlinear differential equation turns into the linear first order differential equation \cite{edwards2000differential}
\begin{align}
\dfrac{dv}{dx}+(1-n)P(x)v=(1-n)Q(x).
\end{align}
\section{Some Necessary Calculations}
\setcounter{equation}{0}
\renewcommand{\theequation}{D\arabic{equation}}
\indent In Section 4.1.1 Equation (16) is given as
\begin{align}
\gamma^{'}(a)+\dfrac{2a[(1+2\omega)\phi^{'2}(a)+\phi(a) \phi^{''}(a)]}{\phi(a) (\phi(a)+a\phi^{'}(a))}\gamma(a)=\dfrac{2(2\omega\rho^{'}a^{3}+3k\phi^{2}(a))}{3a^{3}\phi(a) (\phi(a)+a\phi^{'}(a))}.
\end{align}
For simplification we choose
\begin{align}
\alpha=\phi^{2}+a\phi\phi^{'}.
\end{align}
Then we have
\begin{align}
\phi\phi^{''}=\dfrac{\alpha^{'}-3\phi\phi^{'}-a\phi^{'2}}{a}.
\end{align}
Therefore we can write
\begin{align}
\tilde{P}(a)&=\dfrac{2a[(1+2\omega)\phi^{'2}(a)+\phi(a) \phi^{''}(a)]}{\phi(a) (\phi(a)+a\phi^{'}(a))},\\
&=2\dfrac{\alpha^{'}}{\alpha}+\dfrac{2[2\omega a\phi^{'2}-3\phi\phi^{'}]}{\phi(\phi+a\phi^{'})},\nonumber \\
&=2\dfrac{\alpha^{'}}{\alpha}+P(a),\nonumber
\end{align}
where
\begin{align}
P(a)=\dfrac{2[2\omega a\phi^{'2}-3\phi\phi^{'}]}{\phi(\phi+a\phi^{'})}.
\end{align}
Thus we find
\begin{gather}
\gamma (a)=exp\Big(-\int_{a_{in}}^{a}\tilde{P}(a^{'})da^{'}\Big)\Big\{\int_{a_{in}}^{a}exp\Big(\int_{a_{in}}^{a^{'}}\tilde{P}(a^{''})da^{''}\Big)[\dfrac{2(2\omega\rho^{'}a^{3}+3k\phi^{2}(a))}{3a^{3}\phi(a) (\phi(a)+a\phi^{'}(a))}]da^{'}\\
+\gamma (a_{in})\Big\}.\nonumber
\end{gather}
By using (D.4) we obtain
\begin{align}
exp\Big(-\int_{a_{in}}^{a}\tilde{P}(a^{'})da^{'}\Big)&=exp\Big(-\int_{a_{in}}^{a}[\dfrac{2\alpha^{'}}{\alpha}+P(a^{'})]da^{'}\Big),\\
&=\dfrac{[\phi(a_{in})\Big(\phi(a_{in})+a_{in}\phi^{'}(a_{in})\Big)]^{2}}{[\phi(a)\Big(\phi(a)+a\phi^{'}(a)\Big)]^{2}}exp\Big(-\int_{a_{in}}^{a}P(a^{'})da^{'}\Big).\nonumber
\end{align}
As a result we have
\begin{align}
\gamma(a)&=\dfrac{exp\Big[-\int_{a_{in}}^{a}P(a^{'})da^{'}\Big]}{\Big[\phi (\phi+a\phi^{'})\Big]^{2}}\Big\{\int_{a_{in}}^{a}exp\Big[\int_{a_{in}}^{a^{'}}P(a^{''})da^{''}\Big]Q(a^{'})da^{'}+\tilde{\gamma}(a_{in})\Big\},  \\
P(a)&=\dfrac{2[2\omega a\phi^{'2}(a)-3\phi(a) \phi^{'}(a)]}{\phi(a) (\phi(a)+a\phi^{'}(a))},\nonumber \\Q(a)&=\dfrac{2(2\omega\rho^{'}a^{3}+3k\phi^{2}(a))\Big[\phi(a) (\phi(a)+a\phi^{'}(a))\Big]}{3a^{3}}, \\  \tilde{\gamma}(a_{in})&=\gamma(a_{in})\Big[\phi(a_{in})\Big(\phi(a_{in})+a_{in}\phi^{'}(a_{in})\Big)\Big]^{2}.
\end{align}
\end{appendices}

\bibliographystyle{unsrt}
\bibliography{bd}
\end{document}